\documentstyle[12pt,epsfig]{article}
\begin{document}

\newcommand{\Ktilde}{{K}\kern-.5em\lower 2.75ex\hbox{\Large $\tilde{}$}}
%\newcommand{1\tilde}{{1}\kern-.5em\lower 2.75ex\hbox{\Large 1$\tilde{}$}}

% declarations for front matter

\vspace*{2cm}

\begin{center}
{\Large \bf Effective Field Theory for Pedestrians\footnote{
Invited talk given at the Millennium School on Nuclear and Particle
Physics, National Acceleration Centre, Faure, South Africa,
31 January to 3 February 2000}}\\[1.5cm]
{\large \it G.B. Tupper}\\[.2cm]
Institute of Theoretical Physics and Astrophysics,\\
Department of Physics, University of Cape Town, Rondebosch 7701,\\
South Africa.\\
%{\bf e-mail: viollier@physci.uct.ac.za}
\end{center}

\vspace{2cm}

\begin{abstract}
\noindent
A pedagological introduction to effective field theory is presented.
\end{abstract}

%\maketitle
\newpage

\setlength{\baselineskip}{1.5\baselineskip}
\subsection*{Introduction}
If in 1975 one had asked for a brief history of hadronic physics it would have 
undoubtedly gone something like this [1] : first there was a 
`classical age' initiated by Yuhawa's (1935) meson hypothesis for the nuclear 
force and terminated ($\pm$ 1950) by invading hoards of ``strange'' particles 
and resonance. There followed a sort of `dark ages' where arcane rites of 
dispersion relations, Regge poles and dual resonance models were practiced. 
Finally we are now in the `enlightened age' of ``quantum chromodynamics'' 
(QCD): baryons -- like the proton and neutron -- are composites of three 
``quarks'' while mesons are made of quark-antiquark pairs; these are 
inseparably bound by a colour force which becomes weak at short distances, and 
the interaction between hadrons is a colour analogue of the van der Waal's 
force between neutral atoms.\\[.2cm] 

Alas, some twenty five years later we still are unable to calculate many 
interesting quantities such as the nuclear mass or nucleon-nucleon potential 
directly from QCD (albeit lattice gauge enthusiasts will tell you with the 
next generation of computers $\cdots$). One is left with a variety of models (bag, 
Skyrme, etc.) and a sort of interpolating scheme (QCD sum rules), but nothing 
approaching the systematics and accuracy of quantum electrodynamics (QED). The 
difference is due to confinement: whereas in QED the basic entities (electrons 
and photons) are observable, in QCD they (quarks and gluons) are not, rather 
we can only observe their hadronic composites.\\[.2cm]

Still, the triumphs of QED were afforded by the realization that one did not 
need to be able to calculate the electron mass to determine the effects of the self energy of a bound 
electron -- the Lamb shift [2]. That one could apply a modified version of 
this and work directly with hadrons in a systematic way was first suggested by 
Weinberg (1979) [3] and marked the birth of a new age: the age of effective 
field theory whose ramifications go far beyond hadronic physics alone.\\[.2cm]

There are by now a number of textbook exposition [4] and review articles for 
the sophisticate; in this talk I will endeavour to give the novice some 
feeling for what is going on using the old static model [5] as an 
example. 
Then, at the end I will return to the wider implications.\\[.2cm]

A word of warning: for simplicity (mine, not yours) I will use `natural units'
$\hbar = c$ = 1 ; mass and momenta are in units of energy, and length in units 
of inverse energy, a useful conversion being
%%1
\begin{equation}
\hbar c \; = \; 1 \; = \; 197 \mbox{MeV} \cdot \mbox{fm} \; \; ,
\end{equation} 
(1 fm = 10$^{-13}$ cm) $\; $.\\[.2cm]

\subsection*{The Static Model}
Let me begin by recalling that the impetus for pre-QCD meson theory was Yukawa's observation
that in contrast to Poisson's equation for the electrostatic potential, the equation
\begin{equation}
\left( \frac{\partial^{2}}{\partial t^{2}} - \Delta + m^{2} \right) \; \phi \;
= \; g n
\end{equation}
has for a static charge at the origin, $n (\vec{r}) \; = \; \delta 
(\vec{r})$ 
\begin{equation}
 \phi (r) \; = \; \frac{g}{4 \pi} \;
\frac{e^{-mr}}{r}
\end{equation}
whose range is not infinite but 1/$m$. Now suppose for the moment $n = 0$ ; by making the
the Fourier expansion
\begin{equation}
\phi (t, \vec{r}) \; = \; \int \; \frac{d^{3} k}{(2 \pi^{3})}
\; \varphi \; (t, \vec{k}) \; e^{i \vec{k} \cdot \vec{r}}
\end{equation}
one obtains for each $\vec{k}$
\begin{equation}
\ddot{\varphi} \; (\vec{k}) + \omega^{2} \; (\vec{k}) \;
 \varphi\; (\vec{k}) \; = 0 \; \; , \;
\omega^{2} (\vec{k}) = \vec{k}^{2} \; + \; m^{2} \; \; .
\end{equation}
Thus, classically one has a set of harmonic oscillators and the dispersion relation
$\omega (\vec{k})$ is that for a particle of mass $m$ in relativity.
Each oscillator has a ``momentum''
\begin{equation}
\pi (\vec{k}) \; = \; \dot{\varphi} (\vec{k})
\end{equation}
and the total energy is
\begin{equation}
H_{0} \; = \; \int \; \frac{d^{3} k}{(2 \pi)^{3}} \; \left[
\frac{1}{2} \; \pi^{2} (\vec{k}) + \frac{\omega^{2} (\vec{k})}{2} \; 
\varphi^{2}\;(\vec{k}) \right]
\; \; .
\end{equation}
Now each oscillator can be quantized individually, but instead of $\varphi$
and $\pi$ its more convenient to use $a$ and $a^{+}$
\begin{equation}
\phi \; = \; \frac{1}{\sqrt{2 \omega}} \; (a^{+} + a) \; , \pi = i \;
\sqrt{\frac{\omega}{2}} \; (a^{+} - a) \; \; .
\end{equation}
This gives
\begin{equation}
\hat{H_{0}} \; = \; \int \; \frac{d^{3} k}{(2 \pi)^{3}} \; \omega (k) \;
\hat{a}^{+} (\vec{k}) \; \hat{a} (\vec{k})
\end{equation}
where we have thrown out an infinite sum at ``zero point
 energies''\footnote{Specifically : $\displaystyle{E_{ZPE} \; = \;
\frac{1}{2} \; \int \; \frac{d^{3} k}
 {(2 \pi)^{3}} \; \omega (\vec{k})}$}
 which play no role here. The `ladder operators' 
have non-vanishing commutator
\begin{equation}
\left[ \hat{a} (\vec{k}) \; , \; \hat{a}^{+} (\vec{k}') \right] \; = \; (2 \pi)^{3} \delta \;
 (\vec{k} - \vec{k}\;')
 \end{equation}
 and the lowest energy, ground or `vacuum' state $|0\rangle$ obeys
 \begin{equation}
 \hat{a} (k) |0\rangle \; = \; 0
 \end{equation}
 so indeed it has zero energy. The state $| \vec{k} \rangle = \hat{a}^{+} (\vec{k})
 | 0 \rangle$ has the property
 \begin{equation}
 \hat{H}_{0} \; | \vec{k} \rangle \; = \; \omega (\vec{k}) \; | \vec{k} \rangle
 \end{equation}
 so describing a particle (meson) with (3-) momentum $\vec{k}$ and energy
 $\omega (\vec{k})\;$ .\\[.2cm]

 When the right hand side of (2) is nonzero, i.e. the nucleon is present,
 the oscillators are driven so the total energy (hamiltonian) is
 \begin{equation}
 \hat{H} \; = \; \hat{H}_{0} + \hat{H}_{I} \; \; .
 \end{equation}
 Now, $H_{0}$ is unmodified if the nucleon is static (in practical terms
 this means we are neglecting recoil which is a fair approximation to reality).
 In writing the `interaction part' $H_{I}$ we need to account for the fact
 that the light mesons (pions) are `pseudoscalar'\footnote{In
 QCD this follows from `spontaneously broken chiral symmetry'.}, i.e. under
 `parity', $\vec{r} \rightarrow - \vec{r} \; , \; \phi \rightarrow - \phi$
 whereas for a scalar $\phi$ is unchanged. Taking this together with the
 fact that the nucleon is spin 1/2 (occurring in two spin states, ``up'' and ``down''),
 because the energy should not be changed by parity or rotations the unique
 choice is
 \begin{equation}
 \hat{H}_{I} \; = \; \int \; \frac{d^{3} k}{(2 \pi)^{3}} \;
 \left[ - \frac{i g}{\sqrt{2 \omega (\vec{k})}} \;
 \Ktilde \; \; \left(\hat{a}^{+} (\vec{k}) + \hat{a} (\vec{k})\right) \right]
 \end{equation}
 where $\Ktilde$ $\; \;$ is shorthand for the 2 by 2 matrix
 \begin{eqnarray}
 \Ktilde \; \; = \; \left(
 \begin{array}{ccc}
 k_{z} &\hspace{1cm}& k_{x} - iky\\[.2cm]
 k_{x} + ihy &\hspace{1cm}& - k_{z}\\
 \end{array} \right)
 \end{eqnarray}
 and for simplicity ``isospin'' is neglected. Note the coupling parameter 
$g$ must have dimension of length to compensate that of $\Ktilde \; \;$.\\[.2cm]

 Finally it is also worth mentioning that if one replaces the words nucleon
 and meson by electron and phonon this model bears many similarities to
 problems in solid state physics [6].

 \subsection*{The Self-Energy}
 Alack, unlike $\hat{H}_{0}$, $\hat{H}$ cannot be diagonalized exactly but can be treated
 by time independent perturbation theory familiar to every quantum mechanic.
 Taking the unperturbed state as that with no mesons and one nucleon the leading
 energy shift -- which is to say the nucleon mass shift because it is static
 -- is given by\footnote{This may be compared to the usual expression
%\begin{math}
 $\displaystyle{E_{n}^{(2)} \; = \; \sum_{s \neq n} \; \frac{H_{Ins} \;
 H_{Isn}}{E_{n} - E_{s}}} \; \; $. 
%\end{math}
 Note
$\langle \vec{k} | \hat{H}_{I} |0\rangle = - ig {\Ktilde}
\,\,\sqrt{2 \omega} \;  $.}
 \begin{equation}
 \Delta \; E^{[1]} \; = \; \int \; \frac{d^{3} k}{(2 \pi)^{3}} \;
 \left[ \frac{ig \; \Ktilde \;}{\sqrt{2 \omega}} \right]
 \; \left[ \frac{1}{- \omega} \right] \; \left[
 \frac{- ig \; \Ktilde \;}{\sqrt{2 \omega}} \right]
 \end{equation}
 which can be given a diagrammatic representation\newpage
\begin{figure}[h]
\center
\begin{picture}(0,0)
\epsfig{file=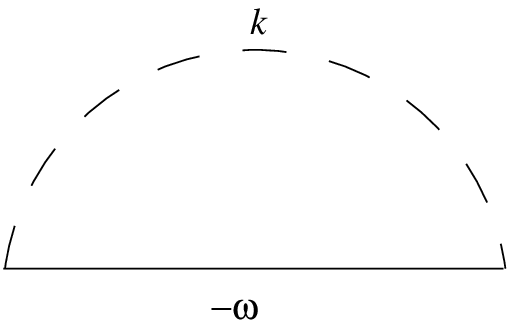}
\end{picture}
\setlength{\unitlength}{3947sp}
\begingroup\makeatletter\ifx\SetFigFont\undefined
\gdef\SetFigFont#1#2#3#4#5{
  \reset@font\fontsize{#1}{#2pt}
  \fontfamily{#3}\fontseries{#4}\fontshape{#5}
  \selectfont}
\fi\endgroup
\begin{picture}(2428,1501)(205,-1350)
\end{picture}
\caption{}
\end{figure}
Reading from right to left : the nucleon emits a meson losing energy
$\omega (\vec{k})$, remains with energy $- \omega$ for a time and then
reabsorbs the meson; it can do this for any $\vec{k}$ so we add all the
intermediate states. Turning this around it is easy to use these `Feynman rules'
to write down contributions corresponding to
\begin{figure}[h]
\center
\begin{picture}(0,0)
\epsfig{file=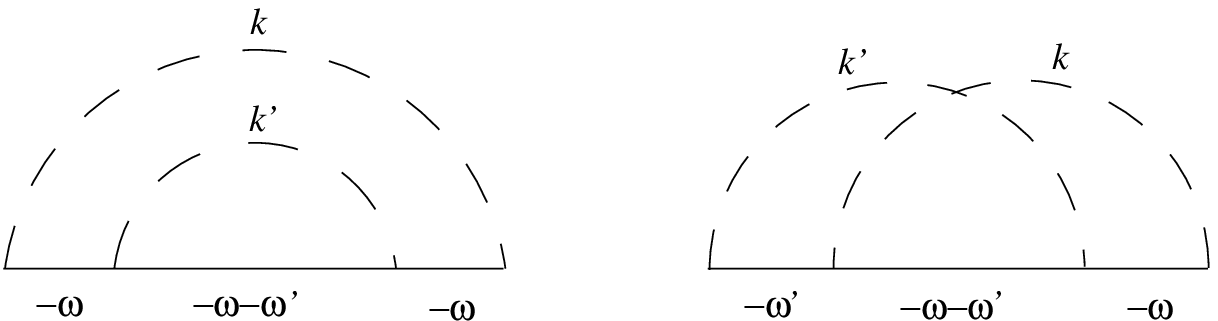}
\end{picture}
\setlength{\unitlength}{3947sp}
\begingroup\makeatletter\ifx\SetFigFont\undefined
\gdef\SetFigFont#1#2#3#4#5{
  \reset@font\fontsize{#1}{#2pt}
  \fontfamily{#3}\fontseries{#4}\fontshape{#5}
  \selectfont}
\fi\endgroup
\begin{picture}(5806,1510)(205,-1359)
\end{picture}
\caption{}
\end{figure}

(try it!). Notice these involve more ``loops''.\\[.2cm]

Of course the energy shift is not a matrix but $(\Ktilde \;)^{2}$ =
$\vec{k}^{2} \;$ I $\;$ and after a little work (16) leads to
\begin{equation}
\delta M^{[1]} \; = \; - \left( \frac{g}{2 \pi} \right)^{2} \; 
\int_{0}^{\infty} dk
\; \left[ k^{2} - m^{2} + \frac{m^{4}}{k^{2} + m^{2}} \right]
\end{equation}
where $k = |\vec{k}|$. It is painfully obvious that only the last integral
converges to $\pi m^{3}/2$, the rest diverge! This is analogous to (even
classical) electrodynamics where in the self-energy of a point change is
infinite.
To be honest we ought to insert a convergence or `form' factor all the
way back in $H_{I}$, but then the result depends on how we choose to 
`cutoff'.\\[.2cm]

Irrespective of details we can say that the nucleon mass $M$ is of the form
\begin{equation}
M \; = \; \hat{M} + \kappa_{1} \; m^{2} - \frac{g^{2}}{8 \pi} \; m^{3} + ....
\end{equation}
where $\hat{M}$, which is what $M$ would be were the meson massless, and
$\kappa_{1}$ `renormalised' parameters hiding the strong cutoff dependence. The
ellipsis
represents weakly cutoff dependent parts, higher loops, etc.
The first significant
thing about (18) is that as a function of $m^{2}$, the parameter appearing
in $H$, the unknown parameters appear in the analytic part whereas the
non-analytic
part is calculable. It is not hard to see why: if we tried to expand
(17) in powers
of $m^{2}$ we soon encounter integrals which diverge at the lower limit
only, and these do
not care how we `regularize'. One reason why this is significant is that in QCD
the pseudoscalar mass squared is proportional to the quark mass, $m^{2}
\propto m_{q}$ ; the first two terms in $M$ give the Gell-Mann-Okubo relation
for the barren octet and the equal splitting rule for the decuplet, the last
the correlation to these.\\[.2cm]

But there is something deeper: the theory we are working with is
`non-renormalizable', signalled by needing $\kappa_{1}$ as a parameter
in the 1-loop calculation. At 2-loops we need more,
and ultimately to hide all our ignorance would require an infinite number of parameters!
Once more, with feeling this time, the bits which are cutoff sensitive
are analytic in $m^{2}$ so
\begin{equation}
M \; = \; \hat{M} + \kappa_{1} \; m^{2} + \kappa_{2} \; m^{4} + \cdots +
\; \mbox{calculable} \; \; .
\end{equation}
\vspace{.2cm}
Now if we replace the upper limit in (17) by $\Lambda$ with
\begin{equation}
\Lambda \; = \; 2 \pi/g
\end{equation}
our one-loop calculation says $K_{1}^{[1]} = \Lambda^{-1}$. Generally then
\begin{equation}
m \; = \; \hat{m} + \bar{\kappa}_{1} \; m^{2}/\Lambda + \bar{\kappa}_{2} \; m^{4}/
\Lambda^{3} + \cdots + \; \mbox{calculable}
\end{equation}
with $\bar{\kappa}_{i}$ a pure number of order unity.\\[.2cm]

We have arrived at the crux of why field theory is effective in the
usual sense of the word. The infinity of parameters do not contribute
equally, and higher orders are suppressed by powers of
$m/\Lambda$.\footnote{Similarly, heavy particle contributions are
suppressed by powers of 1/$m_{H}$. They are subsumed in $\hat{M}$ and
$\bar{\kappa}_{i} \; $.}
Were we calculating meson-nucleon scattering the corresponding
series would be in $|\vec{q}|/\Lambda$ and $m/\Lambda$, $\vec{q}$ the
meson momentum, so this only works for energies low compared to
$\Lambda$. For the case in hand, $m/\Lambda \approx m_{\pi}/m_{\rho}$
$\approx$ 140 MeV/770 MeV and (18) is valid up to the 20\% level (the same
as recoil corrections).

\subsection*{More Effective Theory}
In conclusion, let me stress that our modest calculation did not require
that we know anything about the underlying theory, QCD. All we needed were
the low energy degrees of freedom and their interaction. Now, quantum gravity
is discarded as a fundamental theory because it is nonrenormalizable, involving
as it does the dimensionful newtonian coupling
\begin{equation}
G \; = \; \ell_{p \ell}^{2}
\end{equation}
where $\ell_{p \ell} \approx 10^{-33}$ cm is the Planck length.
As noted by Donoghue [7], however, whatever the ultimate `Theory of
Everything' (GOD) quantum gravity can be treated as an effective field
theory and e.g. quantum corrections to the newtonian potential
\begin{equation}
V(r) \; = \; - \frac{G m_{1} m_{2}}{r} \; \left[ 1 + \beta \left(
\frac{\ell_{p \ell}}{r} \right)^{2} + \cdots \right]
\end{equation}
are calculable. $\beta$ is a computable number of order unity, and
the pathetic smallness of the correction is less significant than
the realization that it can be done.
\newpage
\begin{center}
{\large \bf Appendix}
\end{center}
\vspace{1cm}
In case the reader did try and wants to check his/her work, the expressions
corresponding to figure 2 are
\begin{eqnarray*}
\Delta E^{[2a]} \; = \; \int \frac{d^{3} k}{(2 \pi)^{3}} \; \int \;
               \frac{d^{3} k'}{(2 \pi)^{3}} \left[ \frac{ig\;\Ktilde \;}
               {\sqrt{2 \omega}} \right] \;
 \left[ \frac{1}{- \omega} \right]  \;
        \left[ \frac{ig\;\Ktilde\;\;' \;}{\sqrt{2 \omega'}} \right] \cdot 
\nonumber\\[.5cm] 
            \left[ \frac{1}{- \omega - \omega'} \right] \;
                \left[ \frac{- ig\;\Ktilde\;\;' \;}{\sqrt{2 \omega'}} \right] \;
                \left[ \frac{1}{- \omega} \right] \;
                \left[ \frac{- ig\;\Ktilde \;}{\sqrt{2 \omega}} \right]
\end{eqnarray*}
\begin{eqnarray*}
\Delta E^{[2b]} \; = \; \int \frac{d^{3} k}{(2 \pi)^{3}} \; \int \;
               \frac{d^{3} k'}{(2 \pi)^{3}} \left[ \frac{ig\;\Ktilde\;\;' \;}
               {\sqrt{2 \omega'}} \right] \;
 \left[ \frac{1}{- \omega'} \right]  \;
        \left[ \frac{ig\;\Ktilde \;}{\sqrt{2 \omega}} \right] \cdot 
\nonumber\\[.5cm]
                \left[ \frac{1}{- \omega - \omega'} \right] \;
                \left[ \frac{- ig\;\Ktilde\;\;' \;}{\sqrt{2 \omega'}} \right] \;
                \left[ \frac{1}{- \omega} \right] \;
                \left[ \frac{- ig\;\Ktilde \;}{\sqrt{2 \omega}} \right]
\end{eqnarray*}
These are most difficult to evaluate, but as noted in the text contribute
only at the 20\% level.
\newpage
%\begin{center}
%{\large \bf References}
%\end{center}
%\vspace{1cm}

\end{document}